\documentclass[article,prl,balancelastpage,twocolumn,showpacs]{revtex4}
\usepackage{makeidx}
\usepackage{amssymb}
\usepackage{dcolumn}
\usepackage{graphicx}
\usepackage{acronym}

\input{tcilatex}

\begin{document}

\title{3D transformation for rotating frames and temporal behavior of spin
in rotating electromagnetic fields}
\author{B. V. Gisin }
\affiliation{IPO, Ha-Tannaim St. 9, Tel-Aviv 69209, Israel. E-mail: gisin@eng.tau.ac.il}
\date{\today }

\begin{abstract}
\noindent The Dirac equation is invariant under rotations with a constant
frequency and invariable cylindrical radius. 3D transformation for rotating
frames is found with help of this invariance. Exact localized solutions of
the Dirac equation in the field of a traveling circularly polarized
electromagnetic wave and constant magnetic field exist and possess unusual
properties. Temporal changes of spin pertaining to the solutions are studied.
\end{abstract}

\pacs{03.65.Pm, 03.65.Ta, 31.30.jx, 06.20.Jr\hspace{20cm}}
\maketitle

\section{Introduction}

Usually the transformation to the rotating coordinate system is postulated
without a strong basis like the relativity principle for the Lorentz
transformation. As a rule it is non-Galilean transformation. It means that
time is the same in both frames. A characteristic example is in \cite{Lan}.

In the paper the 3D transformation is deduced using an invariance of Dirac's
equation under rotation with a constant frequency.

We describe properties of exact localized solutions of Dirac equation in
electromagnetic field. We consider temporary evolution of spin by using
spatial averaging. It is well known, the flip-flop of spin is of
considerable interest for the problem of magnetic resonance in quantum
theory \cite{quant}. \ For the exact solutions of the Dirac equation the
temporal behavior of the average spin radically differs from that in the
non-relativistic case.\ \ \ \ \ \ \ \ \ \ \ \ \ \ \ \ \ \ \ \ \ \ \ \ \ \ \
\ \ \ \ \ \ \ \ \ \ \ \ \ \ \ \ \ \ \ \ \ \ \ \ \ \ \ \ \ \ \ \ \ \ \ \ \ \
\ \ \ \ \ \ \ \ \ \ \ \ \ \ \ \ \ \ \ \ \ \ \ \ \ \ \ \ \ 

\section{3D transformation to rotating frames}

\subsection{General form}

Assume that the general form of 3D transformation for cylindrical
coordinates $\varphi ,z,t,$ is%
\begin{eqnarray}
\tilde{\varphi} &=&a_{11}\varphi +a_{12}z+a_{13}t, \\
\tilde{z} &=&a_{21}\varphi +a_{22}z+a_{23}t, \\
\tilde{t} &=&a_{31}\varphi +a_{32}z+a_{33}t,
\end{eqnarray}%
The matrix $a_{nk}$ is a function of the cylindrical radius $r$ and the
frequency of rotation $\Omega $. The cylindrical radius is considered as an
invariable parameter: $\tilde{r}=r.$ The tilde corresponds to the rotating
frame.

We use the dimensionless units 
\begin{equation}
t\rightarrow \frac{ct}{\lambdabar },\text{ \ }(x,y,z)\rightarrow \frac{%
(x,y,z)}{\lambdabar },\text{ \ }\lambdabar =\frac{\hbar }{mc},  \label{nc}
\end{equation}%
$\lambdabar $ is the Compton wavelength.

The transformation must obey the sacral condition of the speed of light
constancy. That is: if $V\equiv z/t=1,$ then $\tilde{V}\equiv \tilde{z}/%
\tilde{t}=1$ for any $\omega \equiv \varphi /t.$ This gives two connections
between coefficients $a_{nk}$%
\begin{equation}
a_{21}=a_{31},\text{ \ }a_{22}+a_{23}=a_{32}+a_{33}.  \label{con}
\end{equation}

Below we determine all the coefficients with help of invariance of Dirac's
equation and using conditions (\ref{con}). Together with the 3D
transformation we define the transformation of spinor to the rotation frame.

\subsection{The Dirac equation}

For construction of the 3D transformation we use the invariance of Dirac's
equation in cylindrical coordinates under rotation with a constant frequency
by invariable cylindrical radius $r$. It means invariance of Dirac's
equation\ without the derivative with respect to $r$.%
\begin{equation}
-i\{\frac{\partial }{\partial t}+\alpha _{1}\frac{1}{2r}+\alpha _{2}\frac{%
\partial }{r\partial \varphi }+\alpha _{3}\frac{\partial }{\partial z}\}\Psi
+\beta \Psi =0.  \label{Dc}
\end{equation}%
Spinor changes as $\Psi \rightarrow \exp (\frac{1}{2}\alpha _{1}\alpha
_{2}\varphi )\Psi $ by translation from the Cartesian $(x,y)$ to cylindrical 
$(x=r\cos \varphi ,$ $y=r\cos \varphi )$ coordinates$.$

An argument in favor such an "invariance by constant cylindrical radius" can
be found in general relativity. Consider Murder's space \cite{Mar}, that is
the solution of Einstein's equation with the cylindrical symmetry with
interval 
\begin{equation}
ds^{2}=Ar^{a+b}dr^{2}+r^{2}d\varphi ^{2}+r^{b}dz^{2}+Cr^{a}dt^{2},
\label{mM}
\end{equation}%
where $A,C,a,b$ are constants. After equating $a=b=2$ and a normalization of 
$r$ and $t$ the metric can be reduced to the form 
\begin{equation}
ds^{2}=(1+\frac{1}{L}r)[dr^{2}+l^{2}d\varphi ^{2}+dz^{2}-c^{2}dt^{2}],
\label{mp}
\end{equation}%
where\emph{\ }$l$ and $L$ are constants with the dimension of length.
Obviously, the interval (\ref{mp}) is invariant under any orthogonal
transformation of coordinates $l\varphi ,$ $z,$ $ict$ and invariable $r.$

\subsection{Invariance}

It is well known that 3D transformation for the Cartesian coordinates can be
composed by rotations around two or third axes. An similar composition can
be used for coordinates $r\varphi ,z,t.$ Consider rotation of the frame
around the $z$-axis with the frequency $\Omega .$ The frequency in the
normalized units $\Omega \rightarrow \Omega \lambdabar /c.$ The Galilean
transformation for this rotation is%
\begin{equation}
\varphi ^{\prime }=\varphi -\Omega t,\text{ \ }t^{\prime }=t  \label{G}
\end{equation}%
From the viewpoint of contemporary physics time in the resting and rotating
frame should be different. Eq. (\ref{Dc}) is not invariant by such a change
of coordinates $\varphi ,t$. The desirable change is a somewhat modified
Lorentz transformations 
\begin{equation}
\varphi ^{\prime }=\frac{\varphi -\Omega t}{\sqrt{1-r^{2}\Omega ^{2}}},\text{
}t^{\prime }=\frac{-r^{2}\Omega \varphi +t}{\sqrt{1-r^{2}\Omega ^{2}}}.
\label{phit}
\end{equation}%
Eq. (\ref{Dc}) is invariant under this transformation. This may be shown by
multiplying Eq. (\ref{Dc}) by the operator%
\begin{equation}
P_{\Phi }=\exp (\frac{1}{2}\alpha _{2}\Phi )  \label{Phi}
\end{equation}%
and the change of spinor $\Psi ^{\prime }=\tilde{P}_{\Phi }\Psi $, where $%
\tilde{P}_{\Phi }=\beta P_{\Phi }\beta =\alpha _{1}P_{\Phi }\alpha _{1}=\exp
(-\frac{1}{2}\alpha _{2}\Phi )$,%
\begin{equation}
\cosh \Phi =\frac{1}{\sqrt{1-r^{2}\Omega ^{2}}},\ \sinh \Phi =\frac{r\Omega 
}{\sqrt{1-r^{2}\Omega ^{2}}}.  \label{chsh}
\end{equation}

The second rotation is in the plane $(r\varphi ,z).$ Eq. (\ref{Dc}) is
invariant under the transformation%
\begin{eqnarray}
r\tilde{\varphi} &=&r\varphi ^{\prime }\cos \Phi _{1}-z\sin \Phi _{1},
\label{phiz} \\
\text{\ }z^{\prime } &=&\varphi ^{\prime }r\sin \Phi _{1}+z\cos \Phi _{1},
\end{eqnarray}%
where $\Phi _{1}$ is a vague angle depending on $r$ and $\Omega $. The wave
function changes as $\Psi ^{\prime \prime }=P_{\Phi 1}\Psi ^{\prime },$where%
\begin{equation}
P_{\Phi 1}=\exp (\frac{1}{2}\alpha _{2}\alpha _{3}\Phi _{1})=\beta \exp (%
\frac{1}{2}\alpha _{2}\alpha _{3}\Phi _{1})\beta  \label{Phi1}
\end{equation}%
The dependence $\Phi _{1}(r,\Omega )$ is defined from the conditions (\ref%
{con}).

The third rotation could be the Lorentz transformation for coordinates $%
z^{\prime }$and $t^{\prime }$ 
\begin{eqnarray}
\tilde{z} &=&z^{\prime }\cosh \Phi _{2}-t^{\prime }\sinh \Phi _{2},
\label{zt} \\
\tilde{t} &=&-z^{\prime }\sinh \Phi _{2}+t^{\prime }\cosh \Phi _{2},
\end{eqnarray}%
where the angle $\Phi _{2}$ is defined by relations%
\begin{equation}
\cosh \Phi _{2}=\frac{1}{\sqrt{1-v^{2}}},\ \sinh \Phi _{2}=\frac{r\Omega }{%
\sqrt{1-v^{2}}},  \label{Phi2}
\end{equation}%
$v$ is the arbitrary normalized velocity $v\rightarrow v/c$. However, for
simplicity, we suppose that $v=0,$ $\Phi _{2}=0.$

With help of conditions (\ref{con}) define the angle $\Phi _{1}$%
\[
\sin \Phi _{1}=-r\Omega ,\text{ \ }\cos \Phi _{1}=\sqrt{1-r^{2}\Omega ^{2}}. 
\]%
Emphasize, for this definition, only one condition from (\ref{con}) is
sufficient, the second is fulfilled automatically.

Using Eqs. (\ref{phit}), (\ref{phiz}) we express $r\tilde{\varphi},$ $\tilde{%
z},$ $\tilde{t}$ as functions of $r\varphi ,$ $z,$ $t.$ The transformation
of coordinates takes the form%
\begin{eqnarray}
\tilde{\varphi} &=&\varphi +z\Omega -\Omega t \\
\tilde{z} &=&\frac{-r^{2}\Omega \varphi }{\sqrt{1-r^{2}\Omega ^{2}}}+z\sqrt{%
1-r^{2}\Omega ^{2}}+\frac{r^{2}\Omega ^{2}t}{\sqrt{1-r^{2}\Omega ^{2}}}, \\
\tilde{t} &=&\frac{-r^{2}\Omega \varphi }{\sqrt{1-r^{2}\Omega ^{2}}}+\frac{t%
}{\sqrt{1-r^{2}\Omega ^{2}}}
\end{eqnarray}%
The determinant of this transformation $|a_{kn}|=1.$

It can be straightforwardly shown that two invariants exist%
\begin{equation}
r^{2}\tilde{\varphi}^{2}+\tilde{z}^{2}-\tilde{t}^{2}=r^{2}\varphi
^{2}+z^{2}-t^{2}  \label{inv1}
\end{equation}%
\begin{equation}
\frac{1}{r^{2}}\frac{\partial ^{2}}{\partial \tilde{\varphi}^{2}}+\frac{%
\partial ^{2}}{\partial \tilde{z}^{2}}-\frac{\partial ^{2}}{\partial \tilde{t%
}^{2}}=\frac{1}{r^{2}}\frac{\partial ^{2}}{\partial \varphi ^{2}}+\frac{%
\partial ^{2}}{\partial z^{2}}-\frac{\partial ^{2}}{\partial t^{2}}
\label{inv2}
\end{equation}%
Note, such an invariance exists by including in the 3D transformation the
third rotation at angle $\Phi _{2}.$

In view of the term $\sqrt{1-r^{2}\Omega ^{2}}$ an upper boundary exists for
the\ radius $r^{2}\Omega ^{2}\leq 1.$ In the non-normalized units this
inequality is equivalent to 
\begin{equation}
\frac{2\pi r}{cT}\leq 1,  \label{rlim}
\end{equation}%
where $T$ is the period corresponding to the frequency $\Omega $. This
obviously means that the speed of points on a circle of radius $r$ cannot be
greater than the speed of light.

It is noteworthy that in the rotation (\ref{phit}) can not arise a new
constant, because then in the inequality (\ref{rlim}) the limit speed would
differ from the speed of light.

Time interval in the rotating and resting frame should be measured at the
same angle and longitudinal coordinate $z$ in corresponding frame. The
dependence of time intervals $\Delta \tilde{t}$ and$\ \Delta t$ is 
\begin{eqnarray}
\Delta \tilde{t} &=&\Delta t\frac{\sqrt{1-\nu ^{2}}\sqrt{1-r^{2}\Omega ^{2}}%
}{1-\nu r^{2}\Omega ^{2}},\text{ \ and\ }  \label{dt1} \\
\Delta \tilde{t} &=&\Delta t\frac{(1-\nu r^{2}\Omega ^{2})}{\sqrt{1-\nu ^{2}}%
\sqrt{1-r^{2}\Omega ^{2}}},  \label{dt2}
\end{eqnarray}%
for $\Delta \tilde{\varphi}=\Delta \tilde{z}=0$ and $\Delta \varphi =\Delta
z=0$ respectively. For the fullness the intervals include rectilinear move
with velocity $v$, that is, the rotation (\ref{zt}).

The operator of the tree dimensional rotations is the product%
\begin{equation}
P=P_{\Phi 1}P_{\Phi }.  \label{Pgen}
\end{equation}%
Define operator $\tilde{P}$%
\begin{equation}
\tilde{P}=\beta P\beta =\exp (\frac{1}{2}\alpha _{2}\alpha _{3}\Phi _{1}-%
\frac{1}{2}\alpha _{2}\Phi ).  \label{Ps}
\end{equation}%
Multiplication by $\tilde{P}$ transforms spinor to the rotating frame $%
\tilde{P}\Psi =\tilde{\Psi}$.

\subsection{Frequency and velocity in the rotating frame}

In accordance with the found transformation, consider the change of
frequency and velocity by transition to the rotating frame 
\begin{eqnarray}
\tilde{\omega} &=&\frac{\omega +v\Omega -\Omega }{-r^{2}\Omega \omega +1}%
\sqrt{1-r^{2}\Omega ^{2}},  \label{f} \\
\tilde{v} &=&\frac{-r^{2}\Omega \omega +v(1-r^{2}\Omega ^{2})+r^{2}\Omega
^{2}}{-r^{2}\Omega \omega +1}  \label{v}
\end{eqnarray}%
At $r^{2}\Omega ^{2}\rightarrow 1$ the frequency $\tilde{\omega}$ tends to
zero and the velocity $\tilde{v}$ along the $\tilde{z}$-axis tends to speed
of light. At $r^{2}\Omega ^{2}\rightarrow 0,$ $\tilde{\omega}\rightarrow
\omega +v\Omega -\Omega ,$ $\tilde{v}\rightarrow v$

\section{The Dirac equation in electromagnetic field}

In this section we study temporal evolution of average spin in rotating
magnetic or electromagnetic fields. This section consist of two part. In the
first part, for comparison, we present non-relativistic case based on the
Pauli equation in rotating magnetic field. The second is devoted to the
Dirac equation in the field of circularly polarized electromagnet wave.

\subsection{Averaging. Nonrelativistic case}

Consider the Pauli equation in the normalized units (\ref{nc}) 
\begin{equation}
i\frac{\partial }{\partial t}\Psi =\frac{1}{2}(\mathbf{p}-\mathbf{A}%
)^{2}\Psi -\frac{g}{2}(\mathbf{\sigma H)}\Psi ,  \label{Pauli}
\end{equation}%
in magnetic field consisting of a constant and circularly polarized part, $g$%
\ is the g-factor. The field is described by the potential $\mathbf{A}$ 
\begin{eqnarray}
A_{x} &=&-H_{z}y/2,\ A_{y}=H_{z}x/2,\text{ }  \label{APx} \\
A_{z} &=&H[-x\sin \Omega t+y\cos \Omega t],  \label{APy}
\end{eqnarray}%
$H_{z}$ is the constant component of the magnetic field along the $z$-axis, $%
H$ is the amplitude of the circularly polarized component. In the normalized
units the potential and magnetic field are%
\begin{equation}
\mathbf{A\rightarrow }\frac{e\lambdabar }{c\hbar }\mathbf{A,}\text{ \ }\frac{%
e\lambdabar ^{2}}{c\hbar }\mathbf{H}\rightarrow \mathbf{H,}  \label{A}
\end{equation}%
The charge, for definiteness, is assumed to be negative $e=-|e|$.

In some cases when the equation for the operator of spin can be separated
from the Pauli equation, the temporal dependence of the spin can be obtained 
\cite{LL}. However the averaging is more general approach.

In experiment, usually, a linearly oscillating magnetic field is used. Such
a field is a combination of two circularly polarized fields with opposite
rotation. Fermion itself chooses the convenient polarization. The second
polarization gains a weak dependence of spin on time, which can be neglected.

The modulation by a circularly polarized magnetic field assumes matching the
polarization of the magnetic field and oscillation spin.

For averaging it is important to assume that solutions are continuous and
square integrable over all the cross-section.

First of all translate spinor to a rotating frame. In this frame the energy
connected with the magnetic moment $g(\mathbf{\sigma H)}/2$ does not depend
on time. As it is known rotation the coordinate system by an angle $\varphi $
corresponds to rotation of spinor by the angle $\varphi /2.$ Rotation the
coordinate system by a frequency $\Omega ,$ or by angle $\Omega t$
corresponds rotation of spinor by the frequency $\Omega /2$. The
transformation is realized by multiplication Eq. (\ref{Pauli}) by $\exp
(i\sigma _{3}\Omega t/2)$ and the change of spinor 
\begin{equation}
\tilde{\Psi}=\exp (\frac{i}{2}\sigma _{3}\Omega t)\Psi .  \label{PsiPr}
\end{equation}

For the Pauli equation, as a non-relativistic equation, the Galilean
transformation coordinates may be used $\tilde{t}=t,$ $\tilde{z}=z,$ $\tilde{%
\varphi}=\varphi -\Omega t,$ $\tilde{x}=r\cos \tilde{\varphi},$ $\tilde{y}%
=r\sin \tilde{\varphi}$. However the same results can be obtained as with as
well as without the coordinate transformation.

Multiply the equation by $\tilde{\Psi}^{\ast }\sigma _{k},$ integrate over
all cross-section and subtract the complex conjugated. We obtain the system
of equations for components of the average spin\ 
\begin{eqnarray}
\frac{\partial }{\partial \tilde{t}}\tilde{s}_{1} &=&(\Omega +gH_{z})\tilde{s%
}_{2},  \label{sp1} \\
\frac{\partial }{\partial \tilde{t}}\tilde{s}_{2} &=&gH\tilde{s}_{3}-(\hbar
\Omega +gH_{z})\tilde{s}_{1},  \label{sp2} \\
\frac{\partial }{\partial \tilde{t}}\tilde{s}_{3} &=&-gH\tilde{s}_{2},
\label{sp3}
\end{eqnarray}%
where $\tilde{s}_{k}=\int \tilde{\Psi}^{\ast }\sigma _{k}\tilde{\Psi}ds$ is
the average spin component with the wave function in the rotating frame, the
integration is over all the cross-section. From this system one follows that
the sum $\tilde{s}_{1}^{2}+\tilde{s}_{2}^{2}+\tilde{s}_{3}^{2}$ do not
depend on time and, without loss generality, it can be normalized to 1.

A solution of this system easily can be found. The required input and output
conditions are: spin oscillations should have a maximum amplitude; the input
and output values of the amplitude should be maximum with differ sign. Using
that we obtain 
\begin{equation}
\tilde{s}_{1}=0,\text{ \ }\tilde{s}_{2}=\sin \omega t,\text{ \ }\tilde{s}%
_{3}=\cos \omega t,\text{ \ }\omega =gH,  \label{sp}
\end{equation}%
provided that the condition of the magnetic resonance\ 
\begin{equation}
\Omega +gH_{z}=0  \label{mres}
\end{equation}%
holds\ \ \ \ \ \ \ \ \ \ \ \ \ \ \ \ \ \ \ \ \ \ \ \ \ \ 

With help of the connection between the wave function in the rotating and
lab frame (\ref{PsiPr}) the spin components with the wave function in the
lab frame $s_{k}=\int \Psi ^{\ast }\sigma _{k}\Psi ds$ also can be found 
\[
s_{1}=-\sin \omega t\sin \Omega t,\text{ }s_{2}=\sin \omega t\cos \Omega t,%
\text{ }s_{3}=\cos \omega t, 
\]

This result is obtained without the concretization of solutions.

In \cite{BVG} on basis of exact solutions of the Pauli equation in the
magnetic field (\ref{APx}), (\ref{APy}), it was shown that the principle of
Pauli should be modified as follows: in such a field, pairs of
non-stationary states exist with the same wave function but with energy
different by $\pm \frac{1}{2}gH$. The sum and difference of the wave
functions corresponds to two states with spins, oscillating as sine and
cosine at the frequency $\omega =gH$.

\subsection{Relativistic case}

In this section unusual properties of exact solutions of the Dirac equation
are described.

Consider Dirac's equation

\begin{equation}
\{-i\frac{\partial }{\partial t}-i\mathbf{\alpha }\frac{\partial }{\partial 
\mathbf{x}}-\mathbf{\alpha \mathbf{A}}+\beta \}\Psi =0.  \label{Dir}
\end{equation}%
in the electromagnetic field with the potential 
\begin{eqnarray}
A_{x} &=&-\frac{1}{2}H_{z}y+\frac{1}{\Omega }H\cos (\Omega t-\Omega z),
\label{Ax} \\
A_{y} &=&\frac{1}{2}H_{z}x+\frac{1}{\Omega }H\sin (\Omega t-\Omega z).
\label{Ay}
\end{eqnarray}%
This potential describes a traveling circularly polarized electromagnetic
wave propagating along constant magnetic field. The normalized coordinates (%
\ref{nc}) as well as the normalized potential are used in the equation (\ref%
{Dir}). In the normalized dimensionless units the propagation constant
equals $\Omega .$

The Dirac's equation (\ref{Dir}) has exact solutions localized in the cross
section perpendicular to the propagation direction of the wave \cite{BVG}.
The solutions in the lab frame can be presented as follows%
\begin{equation}
\Psi =\exp [-iEt+ipz-\frac{1}{2}\alpha _{1}\alpha _{2}(\Omega t-\Omega
z)+D]\psi .  \label{sol0}
\end{equation}%
\begin{equation}
D=-\frac{d}{2}r^{2}-id_{2}\tilde{x}+d_{2}\tilde{y},  \label{D}
\end{equation}%
\begin{equation}
\text{ }d=-\frac{1}{2}H_{z},\text{\ }d_{2}=\frac{\mathcal{E}_{0}h}{2(%
\mathcal{E}-\mathcal{E}_{0})}\frac{1}{\Omega }.  \label{dd2}
\end{equation}%
In the normalized units $d_{2}\rightarrow d_{2}\lambdabar ,$ $d\rightarrow
d\lambdabar ^{2}$ $E\rightarrow E/mc^{2},$ $p\rightarrow p/mc$.

The solution in the rotating frame may be found by means of the Galilean
transformation%
\begin{eqnarray}
\tilde{\varphi} &=&\varphi -\Omega t+\Omega z,\text{ }\tilde{t}=t,\text{ }%
\tilde{z}=z, \\
\tilde{x} &=&r\cos \tilde{\varphi},\text{ \ }\tilde{y}=r\sin \tilde{\varphi},
\\
\tilde{\Psi} &=&\exp [\frac{1}{2}\alpha _{1}\alpha _{2}(\Omega t-\Omega
z)]\Psi .
\end{eqnarray}%
In this section we use the term "rotating frame" bearing in mind "rotating
frame by the Galilean transformation". Using the non-Galilean transformation
in the given case complicates the issue but not leads to new results.

The spinor $\psi $ depend of $\tilde{x},\tilde{y}$. A constant spinor
describe the ground state, a spinor polynomial corresponds to excited states.

In the lab frame only non-stationary states are possible. In contrast to
that in the rotating frame stationary states exist.

$E$ obeys the characteristic equation%
\begin{equation}
\mathcal{E}(\mathcal{E}+2p-\Omega )-1-\frac{\mathcal{E}h^{2}}{\mathcal{E}-%
\mathcal{E}_{0}}=0,  \label{Eqch}
\end{equation}%
\begin{equation}
\text{ }\mathcal{E}_{0}=\frac{2d}{\Omega },\text{ }\mathcal{E}=E-p,\text{\ }%
h=\frac{1}{\Omega }H,  \label{E0}
\end{equation}%
Eq. (\ref{Eqch}) is algebraic equation of the third order. This is one
unusual property of the solutions of the Dirac equation.

The spinor $\psi $ corresponding to the ground state has shape%
\begin{equation}
\psi =N\left( 
\begin{array}{c}
h\mathcal{E} \\ 
-(\mathcal{E}+1)(\mathcal{E}-\mathcal{E}_{0}) \\ 
h\mathcal{E} \\ 
-(\mathcal{E}-1)(\mathcal{E}-\mathcal{E}_{0})%
\end{array}%
\right) ,  \label{solphi}
\end{equation}%
$N$ is the normalization constant, defining from the normalization integral $%
\int \Psi ^{\ast }\Psi ds=1.$%
\begin{equation}
\text{ }N^{2}[h^{2}\mathcal{E}^{2}+(\mathcal{E}^{2}+1)(\mathcal{E}-\mathcal{E%
}_{0})^{2}]\frac{\pi }{d}\exp (\frac{d_{2}^{2}}{d})=1  \label{N}
\end{equation}%
We restrict ourselves the consideration of the ground state.

Spinor (\ref{solphi}) as well as spinors of excited states never can be
presented in the form a small and large two-component spinors. The second
unusual property is the spinors always corresponds only to a relativistic
case.

The parameter $\mathcal{E}_{0}$ in the non-normalized units is defined as%
\begin{equation}
\mathcal{E}_{0}=-\frac{2\mu H_{z}}{\hbar \Omega },  \label{EG}
\end{equation}%
where $\mu =e\hbar /(2mc)$ is the Bohr magneton. Equating $2/\mathcal{E}_{0}$
to g-factor turns the definition (\ref{EG}) in the classical condition of
the magnetic resonance (\ref{mres}).

Evaluate the normalized parameter $h.$ Typically, the amplitude of the
magnetic field $H$ is much smaller than the constant magnetic field $|H_{z}|$
\begin{equation}
h=\frac{1}{\Omega }H\ll -\frac{1}{\Omega }H_{z}=\mathcal{E}_{0}\sim \frac{2}{%
g},  \label{h}
\end{equation}%
If $g$ $\sim 2$ then $h\ll 1.$ After Eq. (\ref{ww}): The typical frequency
of the magnetic resonance $\sim 100GHz.$ The ratio of $\lambda $ to the
Compton wavelength is of the order of $10^{9}.$ Therefore $h$ is extremely
small.

Typically $\mathcal{E}$\ is expanded in power series in $h^{2}.$ However,
for pairs of singular solutions $\mathcal{E}$\ is expanded in power series
in $h$ 
\begin{equation}
\mathcal{E}_{1,2}\mathcal{=E}_{0}+h\mathcal{E}_{1,2}+h^{2}\mathcal{E}%
_{2}+\ldots ,\text{ \ }\mathcal{E}_{1,2}=\pm \frac{\mathcal{E}_{0}}{\sqrt{%
\mathcal{E}_{0}^{2}+1}},  \label{Eh}
\end{equation}%
In this expansion odd terms have positive and negative signs for one and
other solution in the pair. For such pairs in the first approximation%
\begin{equation}
d_{2}\approx \pm \frac{\sqrt{\mathcal{E}_{0}^{2}+1}}{2}.  \label{dpm}
\end{equation}

The necessary condition for existence of the expansion is the equality of
the momentum for both the states in the pair.%
\begin{equation}
p=\frac{1}{\mathcal{E}_{0}}-\mathcal{E}_{0}+\frac{1}{2}\Omega .  \label{p}
\end{equation}%
With this momentum the energy also coincide but with accuracy $\sim h$ 
\begin{equation}
E=\frac{1}{2}(\frac{1}{\mathcal{E}_{0}}+\mathcal{E}_{0})+\frac{1}{2}\Omega
+\ldots  \label{E}
\end{equation}

The pairs could be correspond to the above Pauli pairs. However, the sum and
difference of the wave functions produce states with the spin oscillations
with a vanishingly small amplitude. This amplitude is proportional the
factor $\exp (-2d_{2}^{2}/d).$ In the non-normalized units 
\begin{equation}
\frac{2d_{2}^{2}}{d}=\frac{\mathcal{E}_{0}^{2}+1}{\mathcal{E}_{0}}\frac{%
\lambda }{\lambdabar },  \label{ww}
\end{equation}%
where $\lambda $ is the wavelength corresponding to the frequency $\Omega ,$
The typical frequency of the magnetic resonance $\sim 100GHz,$ the ratio the
wavelength corresponding to this frequency and Compton wavelength is of the
order of $10^{9}.$ Therefore the factor is extremely small

The average value of spin components in the general case are $s_{n}=-\frac{i%
}{2}\int \Psi ^{\ast }\sigma _{n}\Psi ds,$ where 
\begin{equation}
\sigma _{1}=-i\alpha _{2}\alpha _{3},\sigma _{2}=-i\alpha _{3}\alpha
_{1},\sigma _{3}=-i\alpha _{1}\alpha _{2}.
\end{equation}%
The components have the form 
\begin{eqnarray*}
s_{1} &=&\mp \frac{1}{2}\frac{2h\mathcal{E}^{2}(\mathcal{E}-\mathcal{E}_{0})%
}{[h^{2}\mathcal{E}^{2}+(\mathcal{E}^{2}+1)(\mathcal{E}-\mathcal{E}_{0})^{2}]%
}\cos (\Omega t-\Omega z), \\
s_{2} &=&\mp \frac{1}{2}\frac{2h\mathcal{E}^{2}(\mathcal{E}-\mathcal{E}_{0})%
}{[h^{2}\mathcal{E}^{2}+(\mathcal{E}^{2}+1)(\mathcal{E}-\mathcal{E}_{0})^{2}]%
}\sin (\Omega t-\Omega z), \\
s_{3} &=&\frac{1}{2}\frac{h^{2}\mathcal{E}^{2}-(\mathcal{E}^{2}+1)(\mathcal{E%
}-\mathcal{E}_{0})^{2}}{[h^{2}\mathcal{E}^{2}+(\mathcal{E}^{2}+1)(\mathcal{E}%
-\mathcal{E}_{0})^{2}]}.
\end{eqnarray*}

For the singular solutions $\mathcal{E\approx E}_{0}=2/g$ spin $s_{3}=0$ and%
\begin{eqnarray}
s_{1} &=&-\frac{1}{\sqrt{4+g^{2}}}\cos (\Omega t-\Omega z),  \label{s1} \\
\text{\ }s_{2} &=&-\frac{1}{\sqrt{4+g^{2}}}\sin (\Omega t-\Omega z).
\label{s2}
\end{eqnarray}%
The temporal behavior of spin is similar to the magnetic field. This is the
third radical dissimilarity of the solutions.

\subsection{Conclusion}

We have deduced 3D transformation for rotating frames on the basis of the
general assumption regarding its form and the requirement on invariance of \
Dirac's equation under rotation with a constant frequency. The problem
arises by modulation of a rotating magnetic or electromagnetic field.

We have investigated the unusual properties of exact solutions of Dirac's
equation in the field of traveling circularly polarized electromagnetic wave
and a constant magnetic field. The temporal dependence of the average spin
is similar to that of the modulating magnetic field and can be tested by
means of polarization measurements of the spin. It would be interesting to
see how g-factor in these measurements matches known data.

\end{document}